\documentclass{article} 

\usepackage{subfigure}
\usepackage{graphicx} 
\usepackage[english]{babel}
\usepackage{amsmath, amssymb, amsfonts,  bm}
\usepackage{mathrsfs}
\usepackage[mathcal]{euscript}
\usepackage{enumerate}

\usepackage{float}
\usepackage{color,graphicx,epstopdf}
\usepackage{dsfont}
\usepackage{hyperref}
\usepackage{cite} 
\usepackage{algorithm} 
\usepackage{algorithmic} 
\usepackage{tikz-cd}
\usepackage{mathtools}
\usepackage{indentfirst} 
\usepackage{subcaption} 

\usepackage{amsmath}
\usepackage{setspace}
\newtheorem{theorem}{Theorem}
\newtheorem{definition}{Definition}
\newtheorem{lemma}{Lemma}
\newtheorem{remark}{Remark}

\newtheorem{assumption}{Assumption}

\DeclareFontFamily{OT1}{pzc}{}
\DeclareFontShape{OT1}{pzc}{m}{it}{<-> s * [1.200] pzcmi7t}{}
\DeclareMathAlphabet{\mathpzc}{OT1}{pzc}{m}{it}

\newcommand*\mcapinn[2]{\vcenter{\hbox{$\mathsurround=0pt
  \ifx\displaystyle#1\textstyle\else#1\fi\bigcap$}}}

\newcommand*\mcupinn[2]{\vcenter{\hbox{$\mathsurround=0pt
  \ifx\displaystyle#1\textstyle\else#1\fi\bigcup$}}}

\def\begequarr{\begin{eqnarray}}
\def\endequarr{\end{eqnarray}}
\def\begequarrs{\begin{eqnarray*}}
\def\endequarrs{\end{eqnarray*}}
\def\begequ{\begin{equation}}
\def\endequ{\end{equation}}
\def\begequs{\begin{equation*}}
\def\endequs{\end{equation*}}
\def\begite{\begin{itemize}}
\def\endite{\end{itemize}}

\def\begcen{\begin{center}}
\def\endcen{\end{center}}
\def\begrem{\begin{remark}\rm}
\def\endrem{\end{remark}}
\def\ba{\begin{array}}
\def\ea{\end{array}}

\newcommand{\xchc}{\mathbf{\overline{x}}^{\perp}}

\newcommand{\xpic}{\mathbf{\overline{x}}^{\parallel}}

\newcommand{\Ach}{\mathbf{A}^{\perp}}
\newcommand{\AchT}{\mathbf{A}^{\perp T}}
\newcommand{\Api}{\mathbf{A}^{\parallel}}
\newcommand{\ApiT}{\mathbf{A}^{\parallel T}}

\newcommand{\lchc}{\mathbf{\overline{\lamb}}^{\perp}}

\newcommand{\lpic}{\mathbf{\overline{\lamb}}^{\parallel}}

\newcommand{\mV}{\mathrm{V}}

\newcommand{\mE}{\mathrm{E}}

\newcommand{\bb}{\mathbf{b}}

\newcommand{\vb}{\mathbf{v}}

\newcommand{\xb}{\mathbf{x}}
\newcommand{\yb}{\mathbf{y}}
\newcommand{\zb}{\mathbf{z}}

\newcommand{\Lb}{\mathbf{L}}

\newcommand{\Lc}{\mathcal{L}_d}

\newcommand{\Mb}{\mathbf{M}}

\newcommand{\Sbb}{\mathbb{S}}
\newcommand{\Fb}{\mathbf{F}}

\newcommand{\Ebf}{\mathbf{E}}

\newcommand{\Ab}{\mathbf{A}}

\newcommand{\Cb}{\mathbf{C}}
\newcommand{\Cc}{\mathcal{C}}

\newcommand{\Xb}{\mathbf{X}}

\newcommand{\Fl}{\overline{\Fb}}

\newcommand{\xl}{\overline{\xb}}

\newcommand{\xcl}{\overline{\xb}_c}

\newcommand{\nul}{\overline{\bm{\nu}}}

\newcommand{\nub}{\bm{\nu}}
\newcommand{\lamb}{{\bm{\lambda}}}
\newcommand{\laml}{\overline{\bm{\lambda}}}

\newcommand{\sil}{\overline{\boldsymbol{\sigma}}^\xb}
\newcommand{\sill}{\overline{\boldsymbol{\sigma}}^\lamb}

\newcommand{\sigmab}{\boldsymbol{\sigma}}
\newcommand{\onb}{\mathbb{I}}

\title{\LARGE \bf
Linear Convergence  of  Distributed Compressed Optimization with  Equality Constraints
}

\author{Zihao Ren, Lei Wang, Zhengguang Wu, Guodong Shi
\thanks{This research was supported by Zhejiang Provincial Natural Science Foundation of China under Grant No. LZ23F030008, and the National Natural Science Foundation of China under Grant No. 62203386.  Corresponding author: Lei
Wang.}
\thanks{Zihao Ren, Lei Wang and Zhengguang Wu are with the College of Control Science and Engineering, Zhejiang University, P.R. China. (E-mail: zhren2000; lei.wangzju; nashwzhg@zju.edu.cn ).%
}
\thanks{Guodong Shi is with the Australia Centre for Field Robotics, The University of Sydney, Australia (Email: guodong.shi@sydney.edu.au).}
}

\begin{document}

\maketitle
\thispagestyle{empty}
\pagestyle{empty}


\begin{abstract}         
In this paper, the distributed strongly convex optimization problem is studied with spatio-temporal compressed communication and equality constraints. For the case where each agent holds an distributed local equality constraint, a distributed saddle-point algorithm is proposed by employing distributed filters to derive errors of the transmitted states for spatio-temporal compression purposes. It is shown that the resulting distributed compressed algorithm achieves linear convergence. Furthermore, the algorithm is generalized to the case where each agent holds a portion of the global equality constraint, i.e., the constraints across agents are coupled. By introducing an additional design freedom, the global equality constraint is shown to be equivalent to the one where each agent holds an equality constraint, for which the proposed distributed compressed saddle-point algorithm can be adapted to achieve linear convergence.
Numerical simulations are adopted to validate the effectiveness of the proposed algorithms.

\end{abstract}

\section{Introduction}\label{sec.Introduction}
Distributed intelligent systems, such as drone swarms, smart grids, and cyber-physical systems, play an increasingly pivotal role. 
This trend has spurred extensive research across disciplines such as control, signal processing, and machine learning \cite{magnusbook,martinez07,kar12,Rabbat2010}. Distributed optimization is an important problem for a distributed system, which aims to minimize the sum of locally stored functions.
Today,
extensive research has been devoted to studying distributed optimization algorithms, most of which are consensus-based. A simple combination of the consensus algorithm with the gradient descent algorithm results in the distributed subgradient algorithm (DSG), achieving sublinear convergence to the optimal solution \cite{DSMF,DSCO}. To achieve a faster rate of convergence, 
the distributed gradient tracking algorithm (DGT) and the primal-dual algorithm are proposed in \cite{MB-GT} and \cite{XY-LCOF}, respectively, both of which can achieve linear convergence when the objective function is strongly convex. 

The research is also extended for the constrained optimization
case.  For a distributed system involving $n$ agents indexed by $\mathrm{V}=\{1,2,\dots,n\}$, the (in)equality constraints can be divided into two types: distributed constraints, denoted as $g_i(\cdot)\leq0,\forall i\in\mathrm{V}$, and coupled constraints, denoted as $\sum_{i=1}^ng_i(\cdot)\leq 0$. 
For the former, starting from continuous flows, dual variables are introduced to handle the constraints in \cite{AMAS}, which is further extended to discrete algorithms in \cite{DOWM}, where the step size is replaced with a time-varying 
step size. For the latter, in the case of inequalities, \cite{DCOB} and \cite{DSMF2} achieve sublinear convergence based on primal-dual perturbation and Laplace averaging on local copies, respectively.
In addition, interested readers
can refer to \cite{ANCO,DRPA}  and \cite{ODCO,DPDS}
for other constraints cases such as convex set and globally shared constraints. The above research rarely addresses the case for only equality constraints. In this paper, we find that this general case can achieve linear convergence, which has not been discussed by  previous constrained distributed optimization algorithms.


Meanwhile,
in distributed computing, network-based communication between nodes is crucial to the effectiveness of algorithms. \cite{FCRO,CROD} and \cite{FROD,LCOC}, in their respective studies, quantify the communication in DSG and DGT. The research in \cite{XY-CCFD,ALCA,DSOA,AR-AEQD} explores various compressors capable of reducing communication bits by quantization, sparsity, and stochasticity.
In particular, the compressor in \cite{LW-DSFN} incorporates temporal dimensions and the transmitted information becomes a scalar by a time-varying vector. In addition, the communication burden is reduced from the perspective of reducing the number of communication rounds in \cite{evtr}. The common features of compressors in the literature are summarized and a general class of spatio-temporal (ST) compressors are proposed in \cite{ZR-SCCF}, which are applied in this paper.

In this paper, we study distributed strongly convex optimization problems with linear equality constraints. For the case where the constraints are distributed, we propose an algorithm based on saddle-point dynamics. Additionally, we apply a class of ST compressors to the transmitted error states, which are derived by an extra distributed filter. For another case where the equality constraints are coupled, we reformulate an equivalent form of the problem by introducing a new optimization variable. This leads to the development of a saddle-point algorithm with ST compressors for this case as well.
We further prove that both compressed algorithms above can achieve linear convergence.
In numerical simulation, we study the compressed algorithms using several specific compressors belonging to ST compressors for the above two cases to validate the conclusions.
This paper mainly contributes in two aspects to the study of linear equality constrained distributed optimization problems. First, we apply ST compressors, driven by distributed filters, to compress communication over the network. Second, the proposed algorithms achieve linear convergence, thereby advancing the research on convergence rates for constrained distributed optimization.

 

\emph{\bf Notation.}
In this paper, $\|\cdot \|$
denotes the Euclidean norm. The notation $\mathbf{1}_{n}(\mathbf{0}_{n})$, $\mathbf{1}_{n\times d}(\mathbf{0}_{n\times d})$, $\mathbf{I}_n$ and $\{{\bf e}_1,\dots,{\bf e}_m\}$ denote the one (zero) column, the one (zero) matrix, the identity matrix and the base vectors in $\mathbb{R}^d$ respectively. For matrices $\Xb_1,\dots,\Xb_2$ with the same shape,  $\mathrm{blkdiag}(\Xb_1,\dots,\Xb_n)$ is a
diagonal matrix with the $i$-th diagonal matrix being $\Xb_i$. For column vectors $\mathbf{a}$ and $\mathbf{b}$, $[\mathbf{a};\mathbf{b}]$ means $[\mathbf{a}^T,\mathbf{b}^T]^T$.

\section{Problem Formulation}
\label{sec.pro}
\subsection{Constrained Distributed Optimization}

In this paper, we consider a network involving 
$n$ agents. They communicate information through a network described by a graph $\mathrm{G=(V,E)}$, where 
$\mE$ denotes the set of edges. Let $[a_{ij}]\in \mathrm{R}^{n\times n}$ denote the weight matrix complying with graph $\mathrm G$, i.e., $a_{ij}>0$ if $(j,i)\in\mE$ and $a_{ij}=0$ if $(j,i)\notin\mE$.  Denote the neighbor set of agent $i$ as $\mathrm{N}_i$, satisfying $j\in\mathrm{N}_i$ if and only if $a_{ij}\neq0$ for all $i,j\in\mV$. For simplicity, it is assumed
that graph $\mathrm{G}$ is undirected, connected and time-invariant.
Then denote the Laplacian matrix of graph $\mathrm G$ by $\Lb$, satisfying $[\Lb]_{ij}=-a_{ij}$ for all   $i\neq j$, and $[\Lb]_{ii}=\sum_{j=1}^n a_{ij}$ for all $i\in\mathrm{V}$. Denote the neighbor set of agent $i$ as $\mathrm{N}_i$, satisfying $j\in\mathrm{N}_i$ if and only if $[\Lb]_{ij}\neq0$ for all $i,j\in\mV$. This indicates that the Laplacian matrix $\Lb$ is symmetric and positive semi-definite, with $[\Lb]_{ij}=[\Lb]_{ji}$, $ \Lb\mathbf{1}_n=\mathbf{0}_{n}$ and its eigenvalues $\lambda_i$, $i\in\mV$ in an ascending order satisfying $0=\lambda_1<\lambda_2\leq\dots\leq \lambda_n$  by \cite{magnusbook}.  

We are particularly interested in the following two cases of constrained distributed optimization problems. 
\begin{itemize}
    \item [P1)]  
Distributed Linear-equality constrained distributed optimization (DLEC-DO)
    \[
\min_{\xb_i\in\mathbb{R}^d}\sum_{i=1}^n f_i(\xb_i)
    \]
    \begin{equation}
    \label{cdo}
    \ba{c}
    \mathrm{s.t.}\ \xb_i=\xb_j,\  \Ab_i\xb_i-\bb_i=\mathbf{0}, \quad\forall i,j\in \mV.
    \ea
\end{equation}
 \item [P2)]  Coupled-linear-equality constrained distributed optimization (CLEC-DO)
 \[
\min_{\xb_i\in\mathbb{R}^d}\sum_{i=1}^n f_i(\xb_i)
    \]
    \begin{equation}
    \label{cdo2}
    \ba{c}
    \mathrm{s.t.}\ \xb_i=\xb_j,\  \sum_{i=1}^n(\Ab_i\xb_i-\bb_i)=\mathbf{0}, \quad\forall i,j\in \mV,
    \ea
\end{equation}
\end{itemize}
where $f_i:\mathbb{R}^{d}\rightarrow\mathbb{R}$ is Lipschitz continuous, $\Ab_i\in\mathbb{R}^{q\times d}$, $\bb_i\in\mathbb{R}^{q}$.

The first case DLEC-DO is a common case seen in \cite{AMAS,DOWM},  where each agent has a local linear equality constraint. 
For the second case CLEC-DO, the equality constraints are coupled with each other, such as the linear matrix equation in \cite{LW-DSFN}, the distributed Lyapunov equation in \cite{LYA} and the distribution of energy in the electricity system in \cite{MECO}.

 We propose the following assumption about the problem to prepare for the algorithms in this paper.

\begin{assumption}
\label{ass-ec}
   
The function 
$\sum_{i=1}^n f_i(\xb_i)$ is $\mu$-strongly convex. There exist some optimal solutions $\xb_i=\xb^\ast,\ \forall i\in\mathrm{V}$ for DLEC-DO \eqref{cdo} CLEC-DO \eqref{cdo2}, respectively.
\hfill $\square$
\end{assumption}
\subsection{The ST Compressors}

In the work of \cite{ZR-SCCF}, we proposed the ST compressors in discrete time.

\begin{definition}[ST compressor]
\label{def-DST}
$\mathbf{C}:\mathbb{R}^d\times\mathbb{N}_+\rightarrow \mathbb{R}^d$ is a \emph{ST compressor}  if the following properties hold.
\begin{itemize}
    \item [i)] There exists a $L_c>0$ such that $  \|\mathbf{C}(\xb_e,k)\| \leq L_c\|\xb_e\|$ for all $\xb_e\in\mathbb{R}^d$ and any $k\in\mathbb{N}_+$.
    \item [ii)] The discrete-time non-autonomous system $\xb_{e,k+1}=\xb_{e,k}-\kappa_0\mathbf{C}(\xb_{e,k},k)$ is uniformly globally exponentially stable at the origin for some stepsize $\kappa_0>0$.\hfill$\square$
\end{itemize} 
\end{definition}

\begin{remark}
   
In addition to the specific compressors belonging to the ST compressors that we proved in \cite{ZR-SCCF}, we found that the stochastic compressor in \cite{AR-AEQD} and the event-triggered methods in \cite{evtr} can also be classified as ST compressors. The relevant proofs can be extended from \cite{ZR-SCCF} and are omitted in this paper. \hfill$\square$
\end{remark}

\section{Main Results}
\label{sec.mai}

\subsection{ST Compressed Distributed Optimization with Distributed Linear Equality Constraints}

DLEC-DO \eqref{cdo} can be written tightly as 
\[
\min_{\xb\in\mathbb{R}^{nd}}\  \tilde{f}(\xb):=\sum_{i=1}^n f_i(\xb_i)
    \]
\begin{equation}
    \label{cdo_tight}
    \ba{c}
    \mathrm{s.t.}\ \Lc\xb=\mathbf{0},\  \Ab\xb-\bb=\mathbf{0}, 
    \ea
\end{equation}
where
$\Lc:=\Lb\otimes \mathbf{I}_d$, $\xb:=[\xb_{1};\dots;\xb_{n}]$, $\mathbf{A}:=\mathrm{blkdiag}\{\mathbf{A}_1,\dots,\mathbf{A}_n\}$, $\mathbf{b}:=[\mathbf{b}_1;\dots;\mathbf{b}_n]$.

For Problem \eqref{cdo_tight}, we can define a Lagrangian augmented function as 
\[
\ba{rcl}
\mathbb{L}(\xb,\nub,\lamb)&=&\eta\tilde{f}(\xb)+\lamb^T\Lc\xb+\frac{1}{2}\xb^T\Lc\xb+\nub^T(\Ab\xb-\bb)\\&&+\frac{1}{2}\|\Ab\xb-\bb\|^2.
\ea
\]

According to \cite{CO}, the optimal solution with its dual vector is the saddle-point of $\mathbb{L}$. Therefore, the continuous flow in the form of \cite{AMAS} can be obtained by applying the saddle-point dynamic method to $\mathbb{L}$.
 We apply the
ST compressors to the transmitted vector $\xb$ and $\lamb$  with error state compression based on the distributed filter. Finally, we rewrite the algorithm in the distributed form and obtain the following ST Compressed Distributed Optimization with Distributed Linear Equality Constraints (CDC-DE)
algorithm.
\begin{equation}  \label{eq:DE}
\ba{rcl}
\xb_{i,k+1}&=&\xb_{i,k}-\kappa(\Lc\xb_{i,c}+\mathbf{A}_i^T\nub_{i,k}+\Lc\lamb_{i,c}+\eta\nabla\Fb(\xb_{i,k})\\
&&+\Ab_i^T\Ab_i\xb_{i,k}-\Ab_i^T\bb_i),\\
\nub_{i,k+1}&=&\nub_{i,k}+\kappa(\Ab_i\xb_{i,k}-\bb_i),\\
\lamb_{i,k+1}&=&\lamb_{i,k}+\kappa\Lc\xb_{i,c},\\
\sigmab^\xb_{i,k+1}&=&\sigmab^\xb_{i,k}+\kappa_0\Cb(\xb_{i,k}-\sigmab^\xb_{i,k},k),\\
\sigmab^\lamb_{i,k+1}&=&\sigmab^\lamb_{i,k}+\kappa_0\Cb(\lamb_{i,k}-\sigmab^\lamb_{i,k},k),\\
\xb_{i,c}&=&\sigmab_{i,k}^\xb+\Cb(\xb_{i,k}-\sigmab_{i,k}^\xb,k),\\
\lamb_{i,c}&=&\sigmab_{i,k}^\lamb+\Cb(\lamb_{i,k}-\sigmab_{i,k}^\lamb,k).
\ea
\end{equation}

In CDC-DE \eqref{eq:DE}, $\xb_{i,k}$ is the state to reach the optimal solution, while $\nub_{i,k}$ and $\lamb_{i,k}$ are the two dual states in the Lagrangian augmented function $\mathbb{L}$. For communication, $\sigmab^\xb_{i,k}$ and $\sigmab^\lamb_{i,k}$ are the distributed filters for $\xb_{i,k}$ and $\lamb_{i,k}$, respectively, with $\xb_{i,c}$ and $\lamb_{i,c}$ being the transmitted states. It should be noted that only the compressed values $\Cb(\cdot)$ are transmitted in the CDC-DE \eqref{eq:DE} processes (see \cite{XY-CCFD}).

We are ready to propose the theorem for CDC-DE \eqref{eq:DE}.

\begin{theorem}
\label{the-DE}    
For DLEC-DO \eqref{cdo}, let Assumption \ref{ass-ec} hold and $\Cb$ be a ST compressor with $\kappa_0$, then for some $\eta,\kappa>0$, the state $\xb_{i,k}$ in CDC-DE \eqref{eq:DE}  satisfies
$\|\xb_{i,k}-\xb^\ast\|^2=\mathcal{O}(\beta^k)$ for some $\beta\in(0,1)$ and some optimal solution $\xb^\ast$.
\hfill $\square$
\end{theorem}

\subsection{ST Compressed Distributed Optimization with Coupled Linear Equality Constraints}
In this subsection, we will develop an algorithm for the coupled linear equality constraints case, aiming to achieve linear convergence. We propose the following lemma.
\begin{lemma}
\label{lem:eq}
    Problem \eqref{eq:CE} is equal to the following problem. 
    \[
\min_{\xb\in\mathbb{R}^{nd},\zb\in\mathbb{R}^{nq}}\  \tilde{f}(\xb):=\sum_{i=1}^n f_i(\xb_i)
    \]
    \begin{equation}
    \label{cdo2_tight}
    \ba{c}
    \mathrm{s.t.}\ \Lc\xb=\mathbf{0},\  \Ab\xb-\bb+\mathcal{L}_q\zb=\mathbf{0}, 
    \ea
\end{equation}
where $\mathcal{L}_q:=\Lb\otimes \mathbf{I}_q$ and $\zb\in\mathbb{R}^{nq}$ is an optimization variable.
\end{lemma}

{\em Proof:}
It can be proven by showing that $\sum_{i=1}^n(\Ab_i\xb_i-\bb_i)=\mathbf{0}$ if and only if $\Ab\xb-\bb+\mathcal{L}_q\zb=\mathbf{0}$ for some $\zb\in\mathbb{R}^{nq}$.
The necessity can be proved by noting that $\onb_q^T(\Ab\xb-\bb)=\mathbf{0}$, where $\onb_q:=\mathbf{1}_n\otimes\mathbf{I}_q$ is the null space of $\mathcal{L}_q$, so $\Ab\xb-\bb$ lies within the solution space of $\mathcal{L}_q$. And sufficiency can be proven by multiplying $\Ab\xb-\bb+\mathcal{L}_q\zb=\mathbf{0}$ by $\onb_q^T$ on the left. Then Lemma \ref{lem:eq} is proven.\hfill$\square$


A Lagrangian augmented function for Problem \eqref{cdo2_tight} is 
\[
\ba{rcl}
\mathbb{L}(\xb,\zb,\nub,\lamb)&=&\eta\tilde{f}(\xb)+\lamb^T\Lc\xb+\nub^T(\Ab\xb-\bb)\\&&+\frac{1}{2}\|\Ab\xb-\bb\|^2+\nub^T\mathcal{L}_q\zb+\frac{1}{2}\xb^T\Lc\xb
\ea
\]

We use the saddle-point method again and apply the ST compressors to 
the algorithm. Then we rewrite it in the distributed form and obtain the following ST Compressed Distributed Optimization with Coupled Linear Equality Constraints (CDC-CE)
algorithm.
\begin{equation}  \label{eq:CE}
\ba{rcl}
\xb_{i,k+1}&=&\xb_{i,k}-\kappa(\sum_{j=1}^n \Lb_{ij}\xb_{j,c}+\mathbf{A}_i^T\nub_{i,k}\\
&&+\sum_{j=1}^n \Lb_{ij}\lamb_{j,c}+\eta\nabla\Fb(\xb_{i,k})\\
&&+\Ab_i^T\Ab_i\xb_{i,k}-\Ab_i^T\bb_i),\\
\nub_{i,k+1}&=&\nub_{i,k}+\kappa(\Ab_i\xb_{i,k}-\bb_i+\sum_{j=1}^n \Lb_{ij}\zb_{j,c}),\\
\lamb_{i,k+1}&=&\lamb_{i,k}+\kappa(\sum_{j=1}^n \Lb_{ij}\xb_{j,c}),\\
\zb_{i,k+1}&=&\zb_{i,k}-\kappa(\sum_{j=1}^n \Lb_{ij}\nub_{j,c})\\
\sigmab^\xb_{i,k+1}&=&\sigmab^\xb_{i,k}+\kappa_0\Cb(\xb_{i,k}-\sigmab^\xb_{i,k},k),\\
\sigmab^\lamb_{i,k+1}&=&\sigmab^\lamb_{i,k}+\kappa_0\Cb(\lamb_{i,k}-\sigmab^\lamb_{i,k},k),\\
\sigmab^\zb_{i,k+1}&=&\sigmab^\zb_{i,k}+\kappa_0\Cb(\zb_{i,k}-\sigmab^\zb_{i,k},k),\\
\sigmab^{\nub}_{i,k+1}&=&\sigmab^{\nub}_{i,k}+\kappa_0\Cb(\vb_{i,k}-\sigmab^{\nub}_{i,k},k),\\
\xb_{i,c}&=&\sigmab_{i,k}^\xb+\Cb(\xb_{i,k}-\sigmab_{i,k}^\xb,k),\\
\lamb_{i,c}&=&\sigmab_{i,k}^\lamb+\Cb(\lamb_{i,k}-\sigmab_{i,k}^\lamb,k),\\
\zb_{i,c}&=&\sigmab_{i,k}^\zb+\Cb(\zb_{i,k}-\sigmab_{i,k}^\zb,k),\\
\vb_{i,c}&=&\sigmab_{i,k}^{\nub}+\Cb(\vb_{i,k}-\sigmab_{i,k}^{\nub},k).\\
\ea
\end{equation}
 
The states in CDC-CE \eqref{eq:CE} are similar to the states in CDC-DE \eqref{eq:DE}, in which $\nub_{i,k}$, $\lamb_{i,k}$ and $\zb_{i,k}$ are the states in $\mathbb{L}$. Besides, $\sigmab^\xb_{i,k}$, $\sigmab^\lamb_{i,k}$, $\sigmab^\zb_{i,k}$ , $\sigmab_{i,k}^{\nub}$ are the distributed filters and $\xb_{i,c},
\lamb_{i,c},
\zb_{i,c},
\vb_{i,c}$ are the transmitted states.

We are ready to propose the following theorem for CDC-CE \eqref{eq:CE}.

\begin{theorem}
\label{the-CE}    
For CLEC-DO \eqref{cdo2}, let Assumption \ref{ass-ec} hold and $\Cb$ be a ST compressor with $\kappa_0$, then for some $\eta,\kappa>0$, the state $\xb_{i,k}$ in CDC-CE \eqref{eq:CE} satisfies
$  \|\xb_{i,k}-\xb^\ast\|^2=\mathcal{O}(\beta^k)$ for some $\beta\in(0,1)$ and some optimal solution $\xb^\ast$.
\hfill $\square$
\end{theorem}

\section{Conclusion}
\label{sec.con}
In this paper, we have proposed the algorithms for the constrained distributed strongly convex optimization problem, where communication is compressed using ST compressors. For the case where the constraints are distributed linear equalities, a saddle-point algorithm has been proposed. For another case where the equalities are coupled, by studying the problem's equivalent form, the saddle-point algorithm has been adaptive. Both algorithms, incorporating communication compression, achieved linear convergence for constrained distributed optimization problems. Future work may explore the integration of constraint conditions and compression methods, aiming to developing compressors specifically designed to solve constrained problems. Additionally, the advantages of different compressors in addressing various optimization problems could be further investigated.

\appendix

\subsection{Proof of Theorem \ref{the-DE}}
\subsubsection{Optimal Solution of Problem (1)}

The following lemma about optimal solution $\xb^\ast$ of the problem can be obtained easily by KKT condition by Assumption \ref{ass-ec}.

\begin{lemma}
    \label{lem:os}$\xb^\ast\in\mathbb{R}^{nd}$ is the optimal solution of Problem \eqref{cdo_tight} if and only if 
    \begin{equation}
    \label{eq:saddle}
\ba{rcl}
\mathbf{A}\xb^\ast &=& \mathbf{b} ,\\
\Lc \xb^\ast&=&\mathbf{0},\\
 \nabla\Fb(\xb^\ast)&=& \mathbf{A}^T\bm{\nu}^\ast+\Lc \bm{\lambda}^\ast,
\ea
\end{equation}
\end{lemma}
for some  $\nub^\ast\in\mathbb{R}^{nq},\lamb^\ast\in\mathbb{R}^{nd}$, where $\nabla \Fb(\xb):=[\nabla f_1(\xb_{1});\dots;\nabla f_n(\xb_{n})]$.\hfill$\square$



\subsubsection{Analyze of Auxiliary Continuous-Time System}
We first analyze the following continuous-time system instead.

For the system
\begin{equation}  \label{eq:DE_NC_C}
\ba{rcl}
\dot{\xb} &=& -\big(\Lc\xb+\mathbf{A}^T \nub+\Lc\lamb+\eta\nabla\Fb(\xb)  
+\Ab^T\Ab\xb-\Ab^T\bb\big),\\
\dot{\nub} &=& \mathbf{A}\xb-\bb,\\
\dot{\lamb}&=&\Lc\xb,
\ea
\end{equation}

It can be easily obtained that there exists some $(\hat{\xb}^\ast;\hat{\nub}^\ast;\hat{\lamb}^\ast)$ satisfying \eqref{eq:saddle} and $\onb^T\hat{\lamb}^\ast=\onb^T\lamb(0)$ by the existence of optimal solution by Assumption \ref{ass-ec}, where $\onb:=\mathbf{1}_n\otimes\mathbf{I}_d$.
 By Lemma \ref{lem:os}, we can conclude that
$\xb(t)=\hat{\xb}^\ast$, $\lamb(t)=\hat{\lamb}^\ast$, $\nub(t)=\hat{\nub}^\ast$ 
is the equilibrium of \eqref{eq:DE_NC_C}. We define the error state $\xl(t):=\xb(t)-\hat{\xb}^\ast$, $\laml(t):=\lamb(t)-\hat{\lamb}^\ast$, 
$\nul(t):=\nub(t)-\hat{\nub}^\ast$.

Taking the time derivative of the state errors along \eqref{eq:DE_NC_C} yields 
\begin{equation}  \label{eq:DE_NC_C1}
\ba{rcl}
\dot{\xl} &=& -\big(\Lc\xl+\mathbf{A}^T \nul+\Lc\laml+\eta\nabla\Fl(\xl)  +\Ab^T\Ab\xl\big),\\
\dot{\nul} &=& \mathbf{A}\xl,\\
\dot{\laml}&=&\Lc\xl,
\ea
\end{equation}
where $\nabla\Fl(\xl):=\nabla\Fb(\xl+\hat{\xb}^\ast)-\nabla\Fb(\hat{\xb}^\ast)$.

We let $\mathbf{S}\in\mathbb{R}^{n\times(n-1)}$ be a matrix whose rows are eigenvectors corresponding to nonzero eigenvalues of 
$\Lb$,  satisfying
\begin{equation}
 \label{eq:SI}
 \ba {l}
\mathbf{S}^T\mathbf{1}_n =\mathbf{0}_{n-1}\,,\quad  \mathbf{I}_n = \mathbf{S}\mathbf{S}^T +\mathbf{1}_n\mathbf{1}_n^T/n.
\ea
\end{equation}

We define $\Sbb:=\mathbf{S}\otimes\mathbf{I}_d$.
By \eqref{eq:SI}, we can decompose $\xl(t)$ and $\laml(t)$ by defining $\xchc(t):=\Sbb^T\xl(t)$, $\xpic(t):=\onb^T\xl(t)$, $\lchc(t):=\Sbb^T\laml(t)$ and  $\lpic(t):=\onb^T\laml(t)$. We can conclude $\lpic(t)=0$ by $\onb^T\hat{\lamb}^\ast=\onb^T\lamb_0$. Then the system \eqref{eq:DE_NC_C1} becomes 
\begin{equation}  \label{eq:DE_NC_C2}
\ba{rcl}
\dot{\xi} 
&=& \Mb
\xi+ \eta  \Ebf,
\ea
\end{equation}
where $\xi:=[\xchc;\xpic;\nul;\lchc]$ $\Mb:=-\begin{bmatrix}
  \Lambda+\AchT\Ach &\AchT\Api &\AchT  & \Lambda  \cr \ApiT\Ach& \eta +\ApiT\Api & \ApiT \cr -\Ach &-\Api \cr -\Lambda
\end{bmatrix} $, $  \Ebf:=-\begin{bmatrix}
   \Sbb^T\nabla\Fl(\xl) & \onb^T\nabla\Fl(\xl)+ \xpic & \mathbf{0}& \mathbf{0}
\end{bmatrix}^T$,  $\Lambda:=\mathrm{diag}\{\lambda_2,\dots,\lambda_n\}$, $\Ach:=\mathbf{A}\Sbb$ and $\Api:=\mathbf{A}\onb$.

For the nominal system of the system \eqref{eq:DE_NC_C2}
\begin{equation}
\label{eq:DE_NC_Cb}  \ba{rcl}  \dot{\xi} &=& \Mb
\xi,
\ea
\end{equation}
we define a Lyaponuv function $V_1(t)=\frac{1}{2}\|\xi\|^2$, then we have
\[
\ba{rcl}
\dot V_1&=&-\lambda_2\|\xchc\|^2-\eta\|\xpic\|^2-\|\AchT\xchc+\ApiT\xpic\|^2.
\ea
\]
By
LaSalle Invariance Principle \cite{Khalil(2002)}, we know the system \eqref{eq:DE_NC_Cb} satisfies $\lim_{t\to \infty}\|\xchc(t)\|=\lim_{t\to \infty}\|\xpic(t)\|=\lim_{t\to \infty}\|\dot{\nul}(t)\|=0$. Substituting it into \eqref{eq:DE_NC_Cb} and we can conclude $\lim_{t\to \infty}\mathbf{A}^T \nul(t)+\Lc\laml(t)=\lim_{t\to \infty}\mathbf{A}^T \dot{\nul}(t)+\Lc\dot{\laml}(t)=\mathbf{0}$. It means the system \eqref{eq:DE_NC_Cb} will converge to some $({\xb}^\ast;{\nub}^\ast;{\lamb}^\ast)$, which still satisfies \eqref{eq:saddle} and $\onb^T{\lamb}^\ast=\onb^T\lamb(0)$. Thus, by Lemma \ref{lem:os}, if we redefine the error state $\xl(t):=\xb(t)-{\xb}^\ast$, $\laml(t):=\lamb(t)-{\lamb}^\ast$, 
$\nul(t):=\nub(t)-{\nub}^\ast$, with the same process, we can conclude the nominal system \eqref{eq:DE_NC_Cb} converges to zero equilibrium exponentially by noticing it is linear and time-invariant. With \cite{Khalil(2002)}, we know there exists a Lyapunov function $V_{e1}:\mathbb{R}^{(2nd+nq)}\rightarrow\mathbb{R}_+$ which satisfies 
\begin{equation}
\ba{l}\label{eq:Ve_1}
c_1\|\xi_e\|^2 \leq V_{e1}(\xi_e)\leq c_2\|\xi_e\|^2,\\
  \frac{\partial V_{e1}}{\partial \xi_e} \Mb\xi_e \leq -c_3\eta\|\xi_e\|^2,\\\
  \|\frac{\partial V_{e1}}{\partial \xi_e}\| \leq c_4 \|\xi_e\|,
\ea
\end{equation}
for some $c_1,c_2,c_3,c_4>0$.
Besides, it can be noticed that 
\begin{equation}
    \label{eq:Esj}
    \ba{rcl}
\| \Ebf\|\leq \eta (3L_F+2\mu_n)(\|\xchc\|+\|\xpic\|)=: L_E(\|\xchc\|+\|\xpic\|),
    \ea
\end{equation}
which is obtained by the fact 
$\|\Fl(\xl_k)\|\leq L_F(\|\xchc_k\|+\|\xpic_k\|), 
$
for some $L_F>0$ derived by that $f_i$ is Lipschitz continuous.

We define a Lyapunov function as $V_c=pV_1+V_{e,1}$ with some $p>0$.
As  $\sum_{i=1}^n f_i$ is $\mu$-strongly convex, we can obtain that
$
-{\xpic}^T\onb^T \Fl(\xl) 
    \leq -\frac{\mu_n}{2} \|\xpic\|^2+\frac{1}{2\mu_n}L_F^2 \|\xchc\|^2,
$
where $\mu_n:=\mu/n$.
Then with \eqref{eq:Esj}, we have
\begin{equation}
    \label{eq:V_c}
\ba{rcl}
\dot{V}_c
&\leq&  -(p\lambda_2-\eta\zeta_1-\frac{p\eta}{2\mu_n}L_F^2)\|\xchc\|^2\\&&-\eta(p\frac{\mu_n}{2}-\zeta_1)\|\xpic\|^2-\frac{c_3\eta}{3}\|\xi\|^2,
\ea
\end{equation}
where $\zeta_1=\frac{3c_4^2(2L_F+\mu_n)^2}{2c_3}>0$.
As we choose $p=\frac{4\zeta_1}{\mu_n}$ and $\eta\leq\frac{p\lambda_2}{2\zeta_1+pL^2_F/\mu_n}$, then $\dot{V}_c$ is negative definite. With $V_1=\|\xi\|^2$ and \eqref{eq:Ve_1} in mind. We can conclude that the system \eqref{eq:DE_NC_C} exponentially converges to the optimal solution $({\xb}^\ast;{\nub}^\ast;{\lamb}^\ast)$ satisfying \eqref{eq:saddle} and $\onb^T{\lamb}^\ast=\onb^T\lamb_0$. 

\subsubsection{Convergence of CDC-DE (\ref{eq:DE})}

Now we continue to analyze the origin system \eqref{eq:DE}.
It can be written tightly as
\begin{equation}  \label{eq:DE_tight}
\ba{rcl}
\sigmab^\xb_{k+1}&=&\sigmab^\xb_{k}+\kappa_0\Cc(\xb_k-\sigmab^\xb_k,k),\\
\sigmab^\lamb_{k+1}&=&\sigmab^\lamb_{k}+\kappa_0\Cc(\lamb_k-\sigmab^\lamb_k,k),\\
\xb_{k+1}&=&\xb_{k}-\kappa(\Lc\xb_k+\mathbf{A}^T\nub_k+\Lc\lamb_k+\eta\nabla\Fb(\xb_k)\\
&&+\Ab^T\Ab\xb_k-\Ab^T\bb)\\&&+\kappa\Lc(\xb_k-\xb_c)+\kappa\Lc(\lamb_k-\lamb_c),\\
\nub_{k+1}&=&\nub_{k}+\kappa(\mathbf{A}\xb_k-\mathbf{b}),\\
\lamb_{k+1}&=&\lamb_{k}+\kappa\Lc\xb_k-\kappa\Lc(\xb_k-\xb_c),\\
\xb_c&=&\sigmab_{k}^\xb+\Cc(\xb_k-\sigmab_k^\xb,k),\\
\lamb_c&=&\sigmab_{k}^\lamb+\Cc(\lamb_k-\sigmab_k^\lamb,k),
\ea
\end{equation}
where
$\sigmab_k^\xb:=[\sigmab_{1,k}^\xb;\dots;\sigmab_{n,k}^\xb]$, $\sigmab^\lamb_k:=[\sigmab_{1,k}^{\lamb};\dots;\sigmab_{n,k}^{\lamb}]$, 
$\nub_k:=[\nub_{1,k};\dots;\nub_{n,k}]$, $\lamb_k:=[\lamb_{1,k};\dots;\lamb_{n,k}]$ and $\Cc(\xb_k,k):=[\Cb(\xb_{1,k},k);\dots;\Cb(\xb_{n,k},k)]$.

It can be noticed that for $(\xb^\ast;\nub^\ast;\lamb^\ast)$ in the continuous flow \eqref{eq:DE_NC_C}, we conclude that
$\sigmab_k^\xb=\xb_k=\xb_c=\xb^\ast$, $\sigmab_k^\lamb=\lamb_k=\lamb_c=\lamb^\ast$, $\nub_k=\nub^\ast$ 
is the equilibrium of \eqref{eq:DE_tight}. We define the error state $\sil_k:=\sigmab^\xb_k-\xb^\ast$, $\xl_k:=\xb_k-\xb^\ast$, $\xcl:=\xb_c-\xb^\ast$, $\sill_k:=\sigmab^\lamb_k-\lamb^\ast$,  $\laml_k:=\lamb_k-\lamb^\ast$, 
$\laml_c:=\lamb_c-\lamb^\ast$, 
$\nul_k:=\nub_k-\nub^\ast$.
%
%
We similarly 
define  $\xchc_k:=\Sbb^T\xl_k$, $\xpic_k:=\onb^T\xl_k$, $\lchc_k:=\Sbb^T\laml_k$ and  $\lpic_k:=\onb^T\laml_k$. The fact $\lpic_k=0$ still holds. 
Then
we have
\begin{equation}
\label{eq:DE_NC_0}
\ba {rcl}
\sil_{k+1}&=&\sil_{k}+\kappa_0\Cc(\xl_k-\sil_k,k),\\
\sill_{k+1}&=&\sill_{k}+\kappa_0\Cc(\laml_k-\sill_k,k),\\
\xchc_{k+1}&=&\xchc_k-\kappa[  \Lambda \xchc_k+\AchT\Ach \xchc_k+ \AchT\Api\xpic_k +\\&&\eta\Sbb^T \Fl(\xl_k)+\AchT\nul_k]+\kappa\Sbb^T\Lc(\xl_k-\xl_c)+\\&&\kappa\Sbb^T\Lc(\laml_k-\laml_c),\\
\xpic_{k+1}&=&\xpic_k-\kappa[\eta\onb^T\Fl(\xl_k)+\ApiT\Ach\xchc_k+\ApiT\nul_k],\\
   \nul_{k+1} &=& \nul_{k}+\kappa[\Ach\xchc_k+\Api\xpic_k]+\kappa\Sbb^T\Lc(\xl_k-\xl_{c}),\\ {\lchc_{k+1}}&=&\lchc_k+\kappa[ \Lambda\xchc],\\
\xcl&=&\sil_k+\Cc(\xl_k-\sil_k,k),\\
\laml_c&=&\sill_k+\Cc(\laml_k-\sill_k,k).
    \ea
\end{equation}

By definition of $\mathbf{C}(\xb_e,t)$, it is easy to find the following system linearly converges at the zero equilibrium for $\kappa_0>0$, 
\[
\ba{rcl}
{\yb}_{e,k+1}=\yb_{e,k}-\kappa_0\Cc(\yb_{e,k},k).
\ea  
\]
Then there exists positive constants $C$, $\gamma_D<1$, for any $k$ and $N\in\mathbb{N}_+$, the solution satisfies
\[
  \|\yb_{e,k+N}\|^2\leq C(\|\yb_{e,k}\|^2)\gamma_D^N.
\]
Assume $\phi_k^{k+N}(\yb_{e}(k))$ is the state of the system $\yb_{e,k+1}=\yb_{e,k}-\kappa_0\Cc(\yb_{e,k},k)$ in $k+N$ moment with the state in $t$ moment is $\yb_{e,k}$.
It is easy to verify that 
\[\ba{rcl}
  \|\phi_k^{k+N}(\yb)\|^2&\leq& L_\phi \|\yb\|^2,
\ea
\]
for any $\yb\in\mathbb{R}^{nd}$ and some $L_\phi>0$ by property of compressor $\mathbf{C}$.

We find Lyapunov function $V_{e2,k}(\yb_{e,k},k)=\sum_{j=0}^{N-1}\|\phi_k^{k+j}(\yb_{e,k})\|^2$ satisfies 
\begin{equation}
    \label{eq:DPD.a.Ve}
    \ba{rcl}    &&c_{1D}\|\yb_{e}\|^2\leq    V_{e2,k}\leq   c_{2D}\|\yb_e\|^2,
    \ea
\end{equation}
for $c_{1D}=1,c_{2D}=NL_\phi$.

Moreover, we have
\[
\ba{rcl}
      \Delta V_{e2,k}&=&  \sum_{j=1}^{N}\|\phi_{k+1}^{k+j}(\yb_{e,k+1})\|^2\\&&-  \sum_{j=0}^{N-1}\|\phi_k^{k+j}(\yb_{e,k})\|^2\\
    &=&  \|\yb_{e,k+N}\|^2-\|\yb_{e,k}\|^2\\
    &\leq& -(1-C\gamma_D^{N})\|\yb_{e,k}\|^2
    \leq -c_{3D}\|\yb_{e,k}\|^2.
    \ea
\]
We choose a $N\in\mathbb{N}_+$ large enough and then $c_{3D}=1-C\gamma_D^{N}>0$, i.e.,
\begin{equation}
    \ba{rcl}
    \label{eq:DPD.a.Vec3}
&&  \sum_{j=1}^{N}\|\phi_{k+1}^{k+j}(\yb_{e,k}-\kappa_0\Cc(\yb_{e,k},k))\|^2\\&&-  \sum_{j=0}^{N-1}\|\phi_k^{k+j}(\yb_{e,k})\|^2\leq -c_{3D}\|\yb_{e,k}\|^2.
    \ea
\end{equation}

Besides, 
\begin{equation}
\ba{rcl}\label{eq:DPD.a.fact1}
  \|\yb_e-\kappa_0\Cc(\yb_e,k)\|^2\leq \theta\|\yb_e\|^2,
\ea
\end{equation}
for $\theta=2+2
L_c^2\kappa_0^2\lambda_n^2>0$ by property of $\mathbf{C}$.

We define $V_k(\xi_k)=pV_{1,k}(\xi_k)+V_{e1,k}(\xi_k)+V_{e2,k}(\xl_k-\sil_k,k)+V_{e2,k}(\laml_k-\sill_k,k)$. It can be noticed that $   V_k$ is positive definite by \eqref{eq:DPD.a.Ve}.
Then with \eqref{eq:V_c}, \eqref{eq:DPD.a.Vec3} and \eqref{eq:DPD.a.fact1}, we can conclude that for \eqref{eq:DE_NC_0}, there holds
\[
    \ba{rcl}
      \Delta V_k&\leq&
    -(c_{3D}-\kappa\zeta_2)\|\xl_k-\sil_k\|^2\\&&-(c_{3D}-\kappa\zeta_2)\|\laml_k-\sill_k\|^2
    -\kappa[\frac{c_3\eta}{6}\|\xi_k\|^2]\\&&
    +\zeta_3\kappa^2[\|\xi_k\|^2+\|\xl_k-\sil_k\|^2+\|\laml_k-\sill_k\|^2],
    \ea
\]
where 
$\zeta_2=\frac{3}{4c_3\eta}(\lambda_n\theta+\lambda_n\theta c_4+NL_\phi\theta\|\Mb\|+\eta NL_\phi\theta L_E)$ and $\zeta_3=\mathrm{max}\{(2c_4+p+3)(\|\Mb\|^2+L_E^2),3\lambda_n^2\theta NL_\phi\}>0$. We let
$\kappa_1:=\frac{c_{3D}}{2\zeta_2}$,
$\kappa_2:=\frac{1}{2}\mathrm{min}\{\frac{\zeta_2}{\zeta_3},\frac{c_3\eta}{6\zeta_2}\}$,$\kappa_3:=2/\mathrm{min}\{\frac{c_{3D}}{c_{1D}},\frac{c_3\eta}{3p+6c_1}\}$. We can conclude $  \Delta V_k\leq \beta V_k$ for some $\beta=\frac{1}{2}\kappa\mathrm{min}\{\frac{c_{3D}}{c_{1D}},\frac{c_3\eta}{3p+6c_1}\}\in(0,1)$ if $\kappa\leq\min\{\kappa_1,\kappa_2,\kappa_3\}$. With the definition of $V_k$ and $\xl_k=\xb_k-\xb^\ast$, we derive the mean square of $\xb_{i,k}$ in CDC-DE \eqref{eq:DE} converges linearly to the optimal solution of DLEC-DO \eqref{cdo} with ST compressors.

\subsection{Proof of Theorem \ref{the-CE}}
The proof of Theorem 2 is very similar to that of Theorem 1. Therefore, we will only provide a brief outline. We first note that the optimal solution condition \eqref{eq:saddle} becomes
\begin{equation}
    \label{eq:saddle2}
\ba{rcl}
\mathbf{A}\xb^\ast +\mathcal{L}_q\zb^\ast&=& \mathbf{b} ,\\
\Lc \xb^\ast&=&\mathbf{0},\\
 \nabla\Fb(\xb^\ast)&=& \mathbf{A}^T\bm{\nu}^\ast+\Lc \bm{\lambda}^\ast,\\
 \mathcal{L}_q\vb^\ast&=&\mathbf{0},
\ea
\end{equation}
for some  $\lamb^\ast\in\mathbb{R}^{nd},\nub^\ast,\zb^\ast\in\mathbb{R}^{nq}$.
We then introduce the auxiliary continuous-time system similar to \eqref{eq:DE_NC_C}. It can be proved that the system exponentially converges to some $({\xb}^\ast;{\nub}^\ast;{\lamb}^\ast;\zb^\ast)$ satisfying \eqref{eq:saddle2}, $\onb^T{\lamb}^\ast=\onb^T\lamb(0)$ and $\onb_q^T{\zb}^\ast=\onb_q^T\zb(0)$ for some $\eta>0$. Finally, we can prove that with some 
$\kappa>0$, the mean square of $\xb_{i,k}$ in CDC-CE \eqref{eq:CE} converges to the optimal solution of CLEC-DO \eqref{cdo2} linearly with ST compressors.


\end{document}